\documentclass[preprint,showpacs,preprintnumbers,amsmath,amssymb]{revtex4}

\usepackage{graphicx}
\usepackage{dcolumn}
\usepackage{bm}

\begin{document}

\preprint{APS/123-QED}

\title{Magnonic Crystal with Two-Dimensional Periodicity as a Waveguide for Spin Waves}


\author{Rakesh P. Tiwari and D. Stroud}
\affiliation{%
Department of Physics, Ohio State University, Columbus, OH 43210\\
}%

\date{\today}

\begin{abstract}

We describe a simple method of including dissipation in the spin
wave band structure of a periodic ferromagnetic composite, by solving
the Landau-Lifshitz equation for the magnetization with the Gilbert
damping term.  We use this approach to calculate the band structure
of square and triangular arrays of Ni nanocylinders embedded in an Fe host.
The results show that there are certain bands and special directions
in the Brillouin zone where the spin wave lifetime is increased by more
than an order of magnitude above its average value.  
Thus, it may be possible to generate spin waves in such composites decay especially slowly, and propagate especially large distances, for certain frequencies and directions in ${\bf k}$-space.


\end{abstract}

\maketitle


The existence of a periodic superlattice strongly affects many types of excitations in solids.  For example,
the electronic band structure of a conventional semiconductor or semimetal\cite{tsu}, and the dispersion relations of electromagnetic waves\cite{yablonovitch}, 
elastic waves\cite{mm1,mm2,ms1,ms2}, and spin waves\cite{mk,hp,vv,sa,vv2} are all greatly influenced by a periodic
superlattice potential.  In many cases, such potentials can give rise to new, and even complete, electronic, photonic, elastic, or magnonic band gaps which may have 
important implications for the properties of these materials. These excitations have, by now, been extensively studied numerically and analytically, using a variety of 
methods, and have been probed in many experiments\cite{pnas,slv, zk, polushkin}. 

In the present paper, we consider a particular class of such excitations, namely, spin waves in periodic magnetic materials.  Such magnetic superlattices
are often called magnonic crystals. 
 We go beyond previous work by calculating the {\em spin wave lifetimes} in such materials.  Our most striking finding is that the ``figure of merit'' (FOM) of these spin waves (product of spin wave frequency and lifetime) is strongly dependent on the Bloch wave vector {\bf k}, even though, in our model, 
the same spin waves would have a k-independent FOM in a homogeneous magnetic material.  
This strong k-dependence suggests that magnetization in periodic magnetic materials may be transported most efficiently by spin waves 
propagating along special directions in k-space. 
Possibly this k-dependence could be tested by experiments in which spin waves are launched in particular directions corresponding to the largest FOMs. This spin wave generation could be accomplished using real magnetic fields, or (via the spin torque effect\cite{kisilev}) using spin currents.
Measurements of spin wave lifetimes might be carried out, e.\ g., by neutron spin-echo techniques/cite{bayrakci}.


Our calculations are carried out for an array of infinitely long
circular cylinders made of a ferromagnetic material $A$ embedded in another infinite ferromagnetic material $B$. All the 
cylinders are taken to be parallel to the $\hat{z}$ axis and their intersection with the $xy$ plane forms a two-dimensional periodic lattice. We consider two arrangements of such cylinders: a triangular and a square superlattice.  An external static magnetic field {\bf H}$_0$ is applied parallel to the axis of the cylinders, and both ferromagnets are assumed to be magnetized parallel to {\bf H}$_0$.

The equation of motion for this periodic composite is given by the Landau-Lifshitz-Gilbert (LLG) equation\cite{llg}:
\begin{equation}
 \frac{\partial}{\partial t} {\bf M}({\bf r},t)=\gamma\mu_0{\bf M}({\bf r},t)\times{\bf H}_{\text{eff}}({\bf r},t) + \frac{\alpha}{M_s}({\bf r})\left({\bf M}({\bf r},t)
\times \frac{\partial}{\partial t}{\bf M}({\bf r},t)\right).
\label{llg}
\end{equation}
Here $\gamma$ is the gyromagnetic ratio, which is assumed to be the same in both ferromagnets,
{\bf H}$_{\text{eff}}$ is the effective field acting on the magnetization
{\bf M}({\bf r},t), {\bf r} is the position vector, $\alpha$ is the Gilbert damping parameter and $M_s$ is the spontaneous magnetization. For this inhomogenous composite
{\bf H}$_{\text{eff}}$ can be written as
\begin{equation}
 {\bf H}_{\text{eff}}({\bf r},t) = H_0 \hat{z} + {\bf h}({\bf r},t)+\frac{2}{\mu_0M_s}\left(\nabla\cdot\frac{A}{M_s}\nabla\right){\bf M}({\bf r},t),
\end{equation}
where {\bf h}({\bf r},t) is the dynamic dipolar field and $A$ denotes the exchange constant. 
The last term on the right-hand side of eq.\ (2) denotes the exchange field.
For the two-component composite we consider, the exchange constant, the spontaneous magnetization and the Gilbert damping parameter take the forms
$A({\bf r})= A_{B} + \Theta({\bf r})(A_{A}-A_{B})$, 
$M_s({\bf r}) = M_{s,B} + \Theta({\bf r})(M_{s,A}-M_{s,B})$, and
$\alpha({\bf r}) =  \alpha_B + \Theta({\bf r})(\alpha_A-\alpha_B)$,
where the step function 
$\Theta({\bf r})=1$ if ${\bf r}$ is inside ferromagnet A, and
$\Theta({\bf r}) = 0$ otherwise.

We separate the static and time-dependent parts of the magnetization by writing ${\bf M}({\bf r},t)=M_s\hat{z}+{\bf m}({\bf r},t)$, where ${\bf m}({\bf r},t)={\bf m}({\bf r})
e^{-i\omega t}$ is the time-dependent part of the magnetization. The time-dependent dipolar field 
${\bf h}({\bf r})e^{-i\omega t}$, where ${\bf h}({\bf r})=-\nabla\Psi({\bf r})$ and $\Psi({\bf r})$ is the magnetostatic potential. Since 
$\nabla\cdot({\bf h}({\bf r})+ {\bf m}({\bf r}))=0$,
 the magnetostatic potential $\Psi({\bf r})$ obeys 
$\nabla^2\Psi({\bf r})-\nabla\cdot{\bf m}({\bf r})=0$.

Within the linear-magnon approximation\cite{cottam}, the small terms of second order in ${\bf m}({\bf r})$ and ${\bf h}({\bf r})$ are neglected in
the equation of motion. This is equivalent to setting 
${\bf m}({\bf r})\cdot{\bf \hat{z}}=0$\cite{vohl}. 
Substituting the above equations into eqs.\ (\ref{llg}), we obtain
\begin{eqnarray}
i\Omega m_x({\bf r}) + \nabla\cdot [Q\nabla m_y({\bf r})] - m_y({\bf r})- \frac{M_s}{H_0}\frac{\partial\Psi}{\partial y} + i\Omega\alpha m_y({\bf r})&=& 0, \nonumber \\
i\Omega m_y({\bf r}) - \nabla\cdot [Q\nabla m_x({\bf r})] + m_x({\bf r})+ \frac{M_s}{H_0}\frac{\partial\Psi}{\partial x} - i\Omega\alpha m_x({\bf r})&=& 0,
\end{eqnarray}
where $\Omega=\omega/(|\gamma|\mu_0H_0)$ and 
$Q=2A/(M_s\mu_0H_0)$. 

Next, using the periodicity of $Q$, $M_s$ and $\alpha$ in the xy plane, we can expand these quantities in Fourier series as
$Q(\bf x) \equiv Q(x,y)=\sum_{\bf G}Q({\bf G})e^{i{\bf G}\cdot{\bf x}}$, 
with analogous expressions for
$M_s(\bf x)$ and $\alpha({\bf x})$.
Here {\bf x} and {\bf G} are two-dimensional position and reciprocal lattice vectors in the $xy$ plane.
The vector ${\bf r} = ({\bf x}$, $z$), but none of the above quantities will have any z dependence for the
composite we consider.  
The inverse Fourier transforms are of the form
$Q({\bf G})=\frac{1}{S}\int\int d^2{\bf x}Q({\bf x})e^{-i{\bf G}\cdot{\bf x}}$,
where $S$ is the area of the unit cell; similar expressions hold for
$M_s({\bf G})$ and $\alpha({\bf G})$. 

To calculate the band structure for spin waves propagating in the $xy$ plane, we consider the two-dimensional Bloch vector, {\bf k} and use Bloch's theorem to write
$m_x({\bf x})= e^{i{\bf k}\cdot{\bf x}}\sum_{G}m_{x,{\bf K}}({\bf G})e^{i{\bf G}\cdot{\bf x}}$, 
$m_y({\bf x})= e^{i{\bf k}\cdot{\bf x}}\sum_{G}m_{y,{\bf K}}({\bf G})e^{i{\bf G}\cdot{\bf x}}$, and
$\Psi({\bf x})= e^{i{\bf k}\cdot{\bf x}}\sum_{G}\Psi_{\bf K}({\bf G})e^{i{\bf G}\cdot{\bf x}}$.
After some straightforward algebra, the equations of motion reduce to
\begin{equation}
 i\Omega\sum_{\bf {G^{\prime}}}\tilde{A}({\bf G},{\bf G^{\prime}})\left[\begin{array}{c} m_{x,{\bf K}}({\bf G}) \\ m_{y,{\bf K}}({\bf G}) \end{array}\right]= \sum_{\bf {G^{\prime}}}
\tilde{M}({\bf G},{\bf G^{\prime}})
\left[\begin{array}{c} m_{x,{\bf K}}({\bf G^{\prime}}) \\ m_{y,{\bf K}}({\bf G^{\prime}}) \end{array}\right];
\label{llgpwe}
\end{equation}
the 2$\times$2 matrix 
\begin{equation}
 \tilde{A}({\bf G},{\bf G^{\prime}})=\left[\begin{array}{cc}
        \delta_{{\bf G}{\bf G^{\prime}}} & \alpha({\bf G}-{\bf G^{\prime}}) \\
        -\alpha({\bf G}-{\bf G^{\prime}}) & \delta_{{\bf G}{\bf G^{\prime}}},
       \end{array}
\right]
\end{equation}
where $\delta_{{\bf G}{\bf G^{\prime}}}$ is the Kronecker delta and the four components of the 2$\times$2 matrix $\tilde{M}({\bf G},{\bf G^{\prime}})$ are given by
\begin{eqnarray}
 \tilde{M}({\bf G},{\bf G^{\prime}})_{xx}&=&
        \frac{M_s({\bf G}-{\bf G^{\prime}})}{H_0}\frac{(K_x+G^{\prime}_x)(K_y+G^{\prime}_y)}{({\bf K}+{\bf G^{\prime}})^2}       \nonumber \\
\tilde{M}({\bf G},{\bf G^{\prime}})_{xy}&=&
 \delta_{{\bf G}{\bf G^{\prime}}}+
Q({\bf G}-{\bf G^{\prime}})({\bf K}+{\bf G})\cdot({\bf K} + {\bf G^{\prime}}) + \frac{M_s({\bf G}-{\bf G^{\prime}})}{H_0}\frac{(K_y+G^{\prime}_y)^2}
{({\bf K}+{\bf G^{\prime}})^2} \nonumber \\
\tilde{M}({\bf G},{\bf G^{\prime}})_{yx}&=&
         -\delta_{{\bf G}{\bf G^{\prime}}} - Q({\bf G}-{\bf G^{\prime}})({\bf K}+{\bf G})\cdot({\bf K} + {\bf G^{\prime}}) -
 \frac{M_s({\bf G}-{\bf G^{\prime}})}{H_0}\frac{(K_x+G^{\prime}_x)^2}       
{({\bf K}+{\bf G^{\prime}})^2}  \nonumber \\
 \tilde{M}({\bf G},{\bf G^{\prime}})_{yy}&=&
 -\frac{M_s({\bf G}-{\bf G^{\prime}})}{H_0}\frac{(K_x+G^{\prime}_x)(K_y+G^{\prime}_y)}{({\bf K}+{\bf G^{\prime}})^2} .
\end{eqnarray}
On left-multiplying eq.\ (\ref{llgpwe}) by the inverse of the matrix $\tilde{A}$, we reduce the band structure problem, including Gilbert damping, to that of finding the (complex) eigenvalues of $\tilde{A}^{-1}\tilde{M}$. 
A similar plane wave expansion has been previously used to calculate the
magnonic band structure, for the case of zero damping, by several others
(see, e.\ g., Refs.\ \cite{mk} and \cite{llg}).

We have used this formalism to calculate band structures
for both a triangular Bravais lattice, with basis vectors 
${\bf a}_1 = a\hat{x}$, ${\bf a}_2 = a\left(\frac{1}{2}\hat{x} + \frac{\sqrt{3}}{2}\hat{y}\right)$, and a square Bravais lattice, with ${\bf a}_1 = a\hat{x}$,
${\bf a}_2 =a\hat{y}$, where $a$ is the edge of the magnonic crystal unit
cell.  Since Fourier transforms are available analytically for cylinders
of circular cross section, the band structure is easily calculated in this
plane wave representation.

In order to solve Eq.\ (\ref{llgpwe}), we restrict the sum over ${\bf G^{\prime}}$ to the first 625 reciprocal lattice vectors, which
requires the diagonalization of a 1250$\times$1250 complex matrix.
The resulting eigenvalues of the matrix $\tilde{B}({\bf G},{\bf G^{\prime}})$ are all complex.  For a given ${\bf k}$, the
imaginary part of the eigenvalue for gives the spin wave frequency, while
the real part represents the inverse spin wave lifetime.
We have found that both the frequencies and lifetimes are well converged to within 0.1 \% for
this number of plane waves.

For each eigenvalue, the figure of merit (FOM) mentioned above
is the ratio of the imaginary part to the real part of the
eigenvalue.
If the Gilbert damping parameters
$\alpha_A = \alpha_B$, the FOM would be same for all ${\bf k}$'s and
all bands.
By contrast, when $\alpha_A\ne\alpha_B$ we find that the FOM varies from
band to band and depends strongly on ${\bf k}$.  In particular, the
FOM is particularly large in certain high symmetry directions.
As a result, spin waves will have a longer lifetime
when they are launched at special ${\bf k}$ values and with special frequencies. 

We first consider the case of zero damping.  In the left panel of Fig.\ 1, we plot the band structure of a composite of Fe cylinders
arranged on a triangular lattice and embedded in a Ni host, as calculated at an applied field $\mu_0H_0 = 0.1 T$.  The lattice constant $a = 10$ nm and the Fe filling fraction $f = 0.5$ (f is the area fraction occupied by the cylinders).  
The center-hand panel shows a similar composite, but for Fe cylinders arranged
on a square lattice, again with $f = 0.5$.  The right-hand panel shows
the Brillouin zones of the square and triangular lattices with symmetry points
indicated.  In calculating
the band structure, we use an exchange constant and spontaneous magnetization at room temperature of 8.3 pJ/m and 1.71092 $\times$ 10$^6$ A/m for Fe, and 3.4 pJ/m and 0.485423 $\times$ 10$^6$ A/m for Ni\cite{liu}. 
We have not found band structures for exactly these materials in the literature, but when we carry out analogous calculations for Co cylinders in a Permalloy matrix (not shown), using the plane wave method, we obtain nearly identical results to those found by Vasseur {\it et al}\cite{llg}, who also used a plane wave expansion.

In Fig.\ 2, we show analogous calculations including damping for
a square lattice.  We use the same parameters, magnetic field, and value of $f$ as in Fig.\ 1, except that the Gilbert damping parameters are $\alpha_{Fe}=0.0019$ and $\alpha_{Ni}$=0.064, following Ref.\ \cite{oogane}.  In the left panel, the width of
each cross-hatched region is proportional to the figure of merit (FOM)
for the given band and ${\bf k}$ value.  
The right panel shows the FOM for the fourth lowest spin wave band, as a function
of magnonic crystal wave vector ${\bf k}$, along specified directions in the superlattice (or magnonic crystal) Brillouin zone
(SBZ), and at three different filling fractions $f$.  The inset again shows the SBZ and symmetry points. 
We plot the first nine bands. The scales for the FOM and the real frequencies are different, as indicated. 

In Fig.\ 3, we show the corresponding quantities for a triangular magnonic crystal, again using a superlattice constant  $10$ nm and $f = 0.5$.  The other parameters are the same as in Fig.\ 2, except that now the right hand panel shows the FOM for the third lowest spin wave band. 
In Fig.\ 4, we show how the FOM for the optimal special symmetry points of Figs.\ 2 and 3
and bands depends on the superlattice filling fraction $f$.  Note, in particular,
that the FOM increases strongly near the close-packing values of $f$ for both the square and triangular lattices.  


The most striking feature of these plots is the strong dependence of the FOM on both ${\bf k}$ and band index.
For example, in the square superlattice, the FOM is largest in the fourth band at the symmetry point $M$, and in the triangular superlattice, it is largest for the third band at $K$.  The physics behind these strong maxima in
the FOM is that, in both superlattices, the spin waves at these ${\bf k}$-points propagate mainly
through the Fe host, which is the low-damping component.  This result suggests some possible ways to increase
the FOM even further at these points: if we can arrange that a spin wave propagates entirely through the low-dissipation material, this should give an FOM close to the theoretical maximum, which is that of this material
in its homogeneous form.  Thus, a judicious exploration of different periodic composites made of Fe and Ni, or
other materials, could well lead to an even stronger dependence of spin
lifetime on ${\bf k}$ value.

We should add a few words of caution regarding the ``spin waveguiding effect.''
In principle, a measure of distance traveled by a propagating spin waves is given by the coherence length (or spin wave mean free path)$l_c$ \cite{ks}.  This coherence length, for a given band $n$ at wave vector ${\bf k}$, is defined as 
$l_c ({\bf k},n) =|V^{g}_{{\bf k}n}|/{\gamma}_{{\bf k}n}$, where $V^{g}_{{\bf k}n}$ represents the group velocity
and ${\gamma}_{{\bf k}n}$ represents the imaginary part of the eigenfrequency, i. e., the inverse lifetime.  Since the group velocity may itself depend strongly on n and ${\bf k}$, the behavior of $l_c({\bf k}, n)$ may be quite different from
that of the lifetime.  Nevertheless, we expect that $l_c({\bf k}, n)$, like $\tau({\bf k}, n)$ and the FOM $\gamma_{{\bf k}n}$, will depend strongly on both {\bf k} and n, with
sharp extrema near special symmetry points. Hence
the waveguiding effect is likely to remain when one considers $l_c({\bf k}, n)$ 
rather than $\gamma_{{\bf k}n}$.  A full answer to this question would require
a calculation of $V^g_{{\bf k}n}$ for different ${\bf k}$ and $n$.


Since single crystal Fe and Ni already have some intrinsic anisotropy, one might expect that this anisotropy could be exploited to obtain a strongly n and {\bf k}-dependent FOM even in single crystals.  However, in practice, most
magnetic studies of Fe and Ni are carried out on polycrystalline samples, which no longer have this anisotropy.  The present work provides a possible way of recovering this anisotropy, and even more, by use of a periodic lattice of
inclusions.  

The present work can be generalized in various other ways. For example, if a homogeneous magnetic layer is perturbed by a periodic array of spin torque oscillators, this would generate an artificial magnetic superlattice, because the spin torque would provide another contribution to ${\bf H}_{eff}$.  Another
possibility is to extend the present work to magnonic crystals with {\it three-dimensional} periodicity, though this might be 
an experimental challenge.
The present work could conceivably have applications, e.\ g., in
magnonic circuits which exploit the strong anisotropy in magnon lifetimes
found in the present work.    

In summary, we have calculated the spin wave spectrum of a magnetic superlattice with two-dimensional periodicity,
including for the first time the effects of dissipation.  We find a striking anisotropy of the spin wave figure
of merit, which for typical materials is much larger in certain bands near particular points of symmetry in the
Brillouin zone.  This anisotropy implies that propagating spin waves will have much longer lifetimes at certain
frequencies and in certain directions in k-space , which could be interpreted as a waveguiding effect for these excitations.
We suggest that this anisotropy might be further increased with suitable tuning of the array parameters.



Funding for this research was provided by the Center for Emergent Materials at the Ohio State University, an NSF
MRSEC (Award Number DMR-0820414).

\newpage

\newpage

\begin{figure}[h]
\begin{center}
\includegraphics[scale=1.0,angle=0]{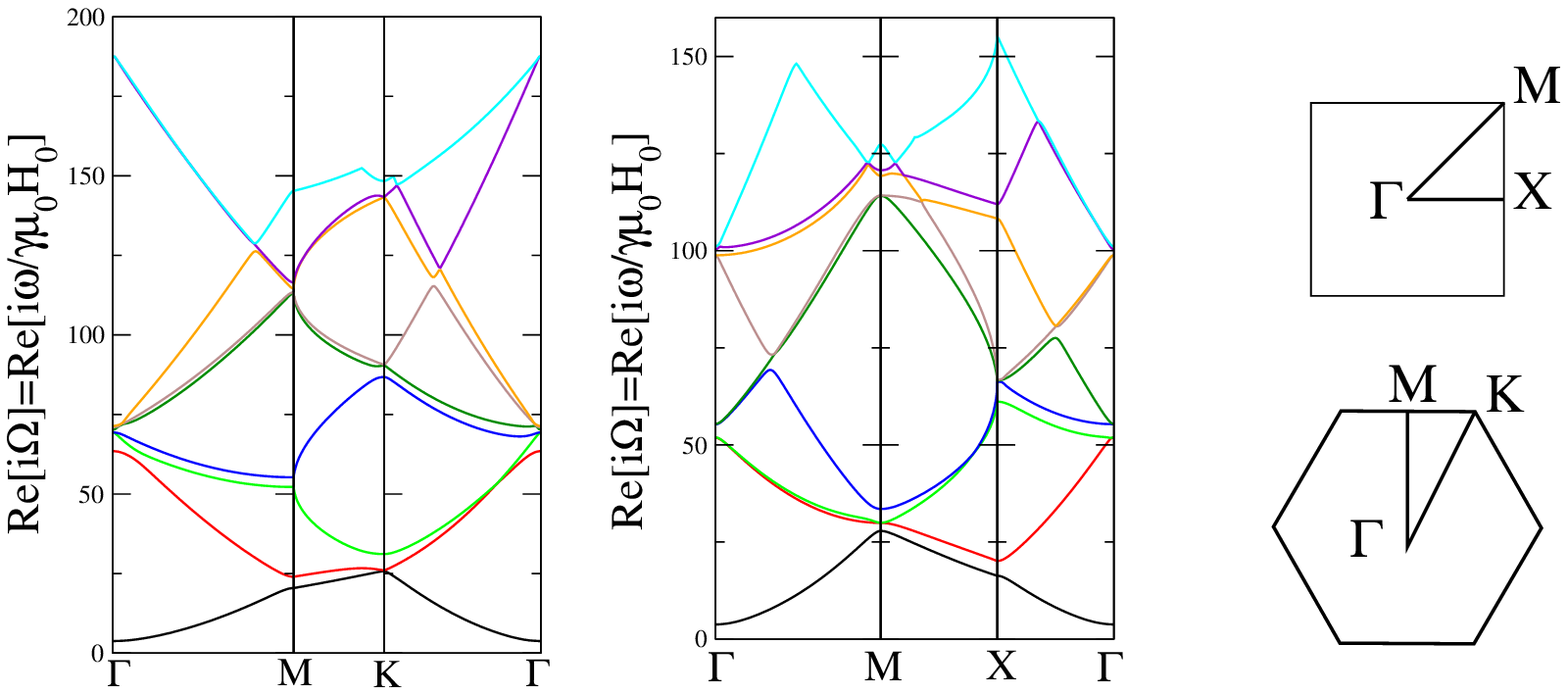}
\caption{(Color online) Left panel: band structure for a triangular lattice of Fe cylinders in Ni, with lattice constant $a = 10$ nm, Fe filling fraction 
$f = 0.5$, and no Gilbert damping.  Other parameters are given in the text.  Center panel: same as left panel but for a square lattice.  Right panel: 
Brillouin zone for square and triangular lattices with symmetry points indicated.}
\end{center}
\end{figure}

\begin{figure}[h]
\begin{center}
\includegraphics[scale=0.5,angle=0]{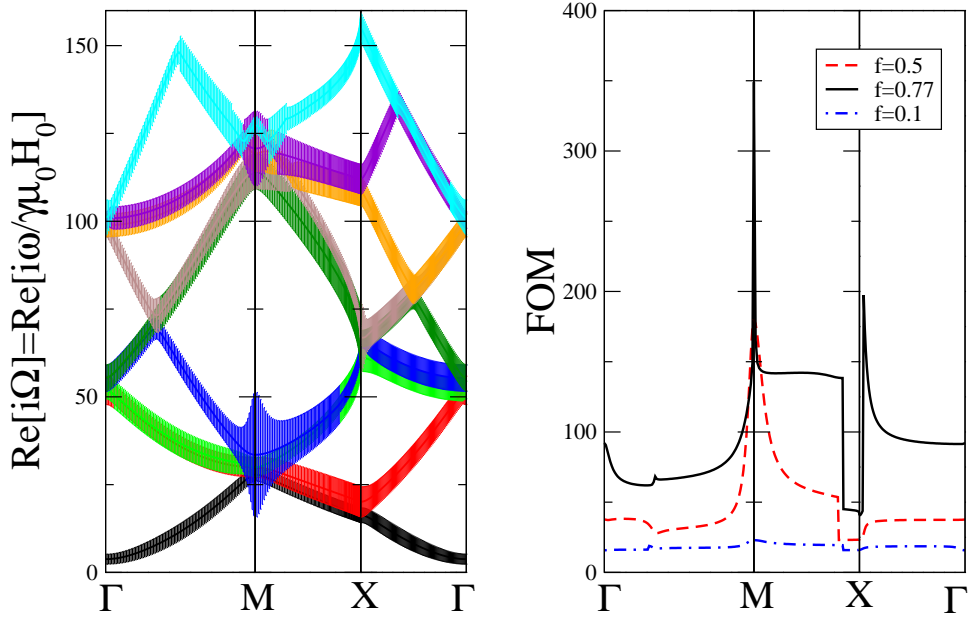}
\caption{(Color Online)  Left panel: same as center panel of Fig.\ 1, but with Gilbert
damping parameters $\alpha_{Fe} = 0.0019$ and $\alpha_{Ni}=0.064$.  The widths of
the cross-hatched regions are proportional to the figure of merit (FOM)
for the given band, as defined in the text.  Right panel: FOM for the fourth lowest spin wave band, as a function
of superlattice wave vector ${\bf k}$, along specified directions in the superlattice Brillouin zone
(SBZ), and at three different filling fractions $f$.}
\end{center}
\end{figure}

\begin{figure}[h]
\begin{center}
\includegraphics[scale=0.5,angle=0]{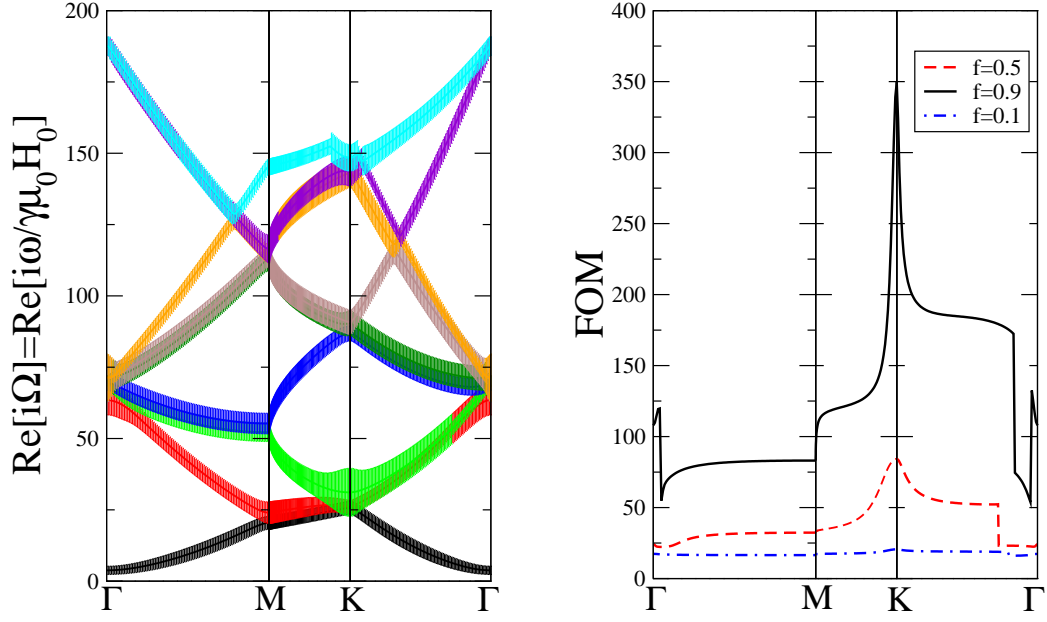}
\caption{(Color online.) Same as Fig.\ 2 but for a triangular lattice
of Fe cylinders in Ni, with lattice constant $a = 10$ and $f = 0.5$ (left panel) and 
$f = 0.1$, $0.5$, and $0.9$ (right panel).}
\end{center}
\end{figure}




\begin{figure}[h]
\begin{center}
\includegraphics[scale=0.5,angle=270]{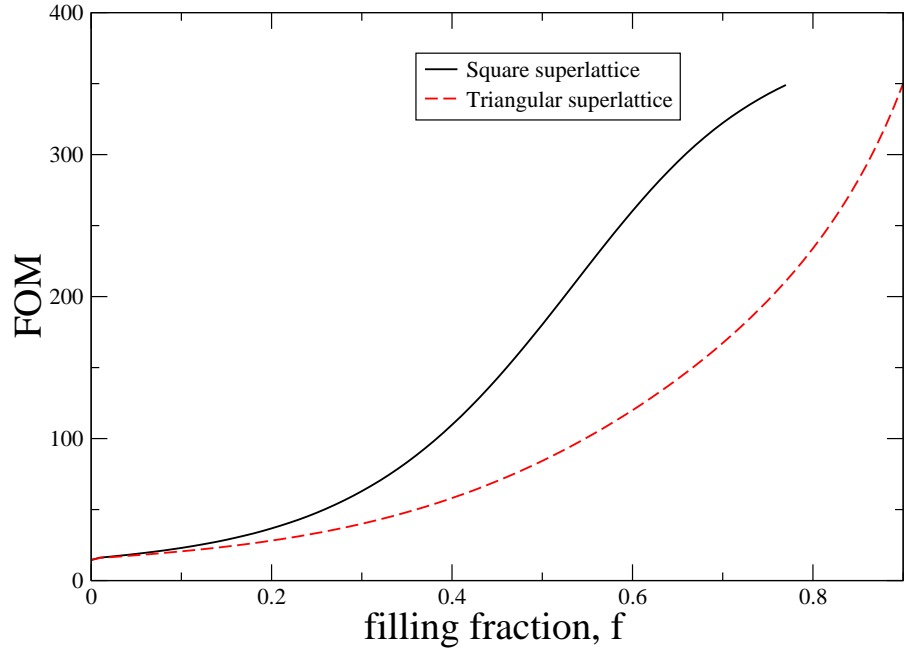}
\caption{(Color Online) Same as Figs.\ 2 and 3, but showing the FOM as a function of filling fraction $f$ for 
$\mu_0{\bf H}_{0} = 0.1 T$, $a = 10$ nm.}
\end{center}
\end{figure}


\end{document}